\documentclass[british,reprint,prd,superscriptaddress,nofootinbib,floatfix]{revtex4-2}
\usepackage[T1]{fontenc}
\usepackage[utf8]{inputenc}
\PassOptionsToPackage{normalem}{ulem}
\usepackage{ulem}
\usepackage{babel}
\usepackage{refstyle}
\usepackage{float}
\usepackage{booktabs}
\usepackage{units}
\usepackage{mathtools}
\usepackage{pdfcomment}
\usepackage{amsmath}
\usepackage{amssymb}
\usepackage{graphicx}
\usepackage{esint}
\ifx\hypersetup\undefined
  \AtBeginDocument{%
    \hypersetup{unicode=true,pdfusetitle,
 bookmarks=true,bookmarksnumbered=true,bookmarksopen=true,bookmarksopenlevel=2,
 breaklinks=false,pdfborder={0 0 0},pdfborderstyle={},backref=false,colorlinks=false}
  }
\else
  \hypersetup{unicode=true,pdfusetitle,
 bookmarks=true,bookmarksnumbered=true,bookmarksopen=true,bookmarksopenlevel=2,
 breaklinks=false,pdfborder={0 0 0},pdfborderstyle={},backref=false,colorlinks=false}
\fi

\makeatletter

\AtBeginDocument{\providecommand\secref[1]{\ref{sec:#1}}}
\AtBeginDocument{\providecommand\figref[1]{\ref{fig:#1}}}
\providecommand{\tabularnewline}{\\}
\RS@ifundefined{subsecref}
  {\newref{subsec}{name = \RSsectxt}}
  {}
\RS@ifundefined{thmref}
  {\def\RSthmtxt{theorem~}\newref{thm}{name = \RSthmtxt}}
  {}
\RS@ifundefined{lemref}
  {\def\RSlemtxt{lemma~}\newref{lem}{name = \RSlemtxt}}
  {}

\usepackage{colortbl}
\usepackage[official]{eurosym}
\usepackage{braket}
\usepackage{bm}
\usepackage{upgreek}
\usepackage{diagbox}
\usepackage{lipsum}
\usepackage{epigraph}
\usepackage{url}

\usepackage{tensind}
\tensordelimiter{?}
\usepackage{pgfplots}
\usepackage{pgfplotstable}
\usetikzlibrary{pgfplots.groupplots}

\DeclareSymbolFont{vectors}{OML}{cmm}{b}{it}
\DeclareSymbolFont{tensors}{OT1}{cmss}{bx}{it}

\DeclareSymbolFontAlphabet{\mathvec} {vectors}
\DeclareSymbolFontAlphabet{\mathtens}{tensors}

\makeatother

\begin{document}
\global\long\def\tudu#1#2#3#4{{?{#1}^{#2}{}_{#3}{}^{#4}?}}%

\global\long\def\tud#1#2#3{{?{#1}^{#2}{}_{#3}?}}%

\global\long\def\tudud#1#2#3#4#5{{?{#1}^{#2}{}_{#3}{}^{#4}{}_{#5}?}}%

\global\long\def\tdu#1#2#3{{?{#1}_{#2}{}^{#3}?}}%

\global\long\def\dd#1#2{\frac{\mathrm{d}#1}{\mathrm{d}#2}}%

\global\long\def\pd#1#2{\frac{\partial#1}{\partial#2}}%

\global\long\def\tens#1{\mathtens{#1}}%

\global\long\def\threevec#1{\mathvec{#1}}%

\global\long\def\d{\mathrm{d}}%

\global\long\def\e{\mathrm{e}}%

\global\long\def\eps{\varepsilon}%

\global\long\def\i{\mathrm{i}}%

\global\long\def\ext{\tilde{\mathrm{d}}}%

\title{Stability and physical properties of spherical excited scalar boson stars}

\author{Marco Brito}
\affiliation{Departamento de Matemática da Universidade de Aveiro \\
and Centre for Research and Development in Mathematics and Applications (CIDMA) \\
Campus de Santiago, 3810-183 Aveiro, Portugal}
\author{Carlos Herdeiro}
\affiliation{Departamento de Matemática da Universidade de Aveiro \\
and Centre for Research and Development in Mathematics and Applications (CIDMA) \\
Campus de Santiago, 3810-183 Aveiro, Portugal}
\author{Eugen Radu}
\affiliation{Departamento de Matemática da Universidade de Aveiro \\
and Centre for Research and Development in Mathematics and Applications (CIDMA) \\
Campus de Santiago, 3810-183 Aveiro, Portugal}
\author{Nicolas Sanchis-Gual}
\affiliation{Departamento de Astronomía y Astrofísica, Universitat de València, \\
Dr.\ Moliner 50, 46100, Burjassot (Valencia), Spain}
\affiliation{Departamento de Matemática da Universidade de Aveiro \\
and Centre for Research and Development in Mathematics and Applications (CIDMA) \\
Campus de Santiago, 3810-183 Aveiro, Portugal}
\author{Miguel Zilhão}
\affiliation{Departamento de Matemática da Universidade de Aveiro \\
and Centre for Research and Development in Mathematics and Applications (CIDMA) \\
Campus de Santiago, 3810-183 Aveiro, Portugal}

\begin{abstract}
We study the time evolution of spherical,  excited -- with $n$ radial nodes -- 
scalar boson stars in General Relativity minimally coupled to a complex massive scalar field with quartic self-interactions. We report
that these stars, with  up to $n=10$, can be made dynamically stable, up to timescales
of $t\sim\frac{10^{4}}{c\mu}$, where $\mu$ is the inverse Compton wavelength of the scalar particle,  
for sufficiently large values of the self-interactions
coupling constant $\lambda$, which depend on $n$. We observe that the compactness of these solutions is rather insensitive to $n$, for large $\lambda$ and fixed frequency. Generically, along the branches where stability was studied, these excited boson stars are not compact enough to allow for innermost stable circular orbits or light rings. Finally, we discuss the angular velocity of particles along timelike circular
orbits, suggesting an application, for solutions in the Newtonian limit, to galactic rotation curves.
\end{abstract}

\maketitle

\section{Introduction}

Bosonic stars are localised self-gravitating lumps of bosonic fields, sustained
by their own gravitational pull and (possibly) self-interactions, depending
on the chosen model -- see~\cite{Schunck:2003kk,Liebling:2012fv} for reviews. These hypothetical stars could be an alternative
to (or co-existing with)  black holes, \emph{e.g.}~\cite{Schunck:1998cdq,Mielke:2000mh,Berti:2006qt,Guzman:2009zz,Vincent:2015xta,Grould:2017rzz,Olivares:2018abq,Herdeiro:2021lwl,Rosa:2022tfv}. Their dynamics can match real  gravitational waves signals~\cite{Bustillo2021,CalderonBustillo:2022cja} and they have long been hypothesised as possible dark matter constituents~\cite{Lee:1995af,Suarez:2013iw,Eby:2015hsq,Chen:2020cef}. In order for such stars to be realisable in the real world, however, 
they must be stable against small perturbations, which are unavoidable in any realistic astrophysical environment.

\medskip

Let us consider spherically symmetric scalar
boson stars, which will be the focus of this work - see \emph{e.g.}~\cite{Colpi:1986ye,Bernal:2009zy,Hartmann:2013tca,Herdeiro:2017fhv,Alcubierre:2018ahf,Brihaye:2018grv,Guerra:2019srj,Boskovic:2021nfs,Maso-Ferrando:2021ngp} for different models. These stars may be found in different states,
ranging from the fundamental or ground state ($n=0$) to an infinitude
of excited states ($n\geq1$), analogous with the orbitals of the
hydrogen atom -- the $(n+1)\mathrm{s}$-orbital. The value of $n$ is the number of nodes of the
radial scalar profile. When a scalar field
with no self-interactions is considered (\emph{i.e.}\ mini boson stars),
excited states are possible, but they are unstable, decaying either
to the fundamental state or collapsing into a black hole~\citep{Balakrishna1997}. Mini boson stars in the ground state, by contrast,  are stable (along the appropriate branch)~\citep{Seidel1990}. 
It was recently reported~\citep{SanchisGual2022}, however, 
that when a quartic self-interaction is introduced, it has a healing power which does stabilise
$n=1$ stars, for high enough values of $\lambda$, the coupling
constant ruling the self-interactions. It may be expected that such healing behaviour could extend to higher values of $n$. To assess and establish this possibility is a  central purpose of this paper.

\medskip

If excited stars are dynamically stable, they may have astrophysical applications. An interesting question is if such stars can be black hole foils. For instance, if they can be compact enough to allow the existence
of special orbits that are characteristic of black holes, namely light rings~\cite{Cunha:2020azh} or an ISCO (innermost
stable circular orbit)~\cite{Delgado:2021jxd}. The lensing properties of ultracompact spherical bosonic stars (\emph{i.e.}\ possessing light rings) has been considered in~\cite{Cunha:2017wao}. But recent evidence suggests such stars may be generically unstable~\cite{Cunha:2017qtt,Cunha:2022gde}. On the other hand, even if they are not ultracompact, a certain structure of the timelike circular geodesics could void the central region of bosonic stars of emitting matter, therefore creating an ``effective'' shadow ~\cite{Olivares:2018abq,Herdeiro:2021lwl}. We will investigate if this can occur for the excited models we consider here. We remark that even if none of these features is present (ISCOs, light rings or the aforementioned special structure of timelike circular geodesics), boson stars could still   mimic black hole data in certain situations, such as the emission of gravitational
waves in particular events~\cite{Bustillo2021,CalderonBustillo:2022cja}.
As another potential astrophysical application, we will consider the possibility that the excited boson stars could fit galactic rotation curves~\citep{Lee:1995af}, and, consequently, whether they could be realistic models for
dark matter galactic halos.

This paper is organised as follows. In \secref{The-model} we describe
the excited states of spherical boson stars, as solutions of the appropriate
Einstein-Klein-Gordon system, discussing their masses and compactness. In \secref{evolutions} we discuss the numerical relativity framework for our evolutions. 
In \secref{Results} we discuss the  main results both in the stability and formation scenarios.
In \secref{ISCO-and-angular} we study timelike geodesics around
boson stars spacetimes, probing if ISCOs or light rings are possible
in any of our boson star models. We also study the structure of timelike circular orbits and a possible connection to galactic rotation curves.
We close with a discussion and final remarks. For the rest of the article we shall use the metric signature $(-,+,+,+)$
and set $c=G=1$.

\bigskip

\section{\label{sec:The-model}The model}

\subsection{The action and field equations}
 We consider the following action describing
Einstein’s gravity minimally coupled to a complex, massive, self-interacting
scalar field $\Phi$ (with complex conjugate $\Phi^{*}$)

\begin{equation}
S=\int \bigg[\frac{R}{16\pi}
-\frac{1}{2} (\Phi_{,\mu}^{*}\Phi^{,\mu}
+U(|\Phi|^{2}) ) \bigg]
\sqrt{-g}\,\d^{4}x
\end{equation}
which results in the following field equations
\begin{equation}
R_{\mu\nu}-\frac{1}{2}g_{\mu\nu}R=8\pi T_{\mu\nu},\label{eq:einstein-eq}
\end{equation}
\begin{equation}
\square\Phi=\dd U{|\Phi|^{2}}\Phi,\label{eq:k-g}
\end{equation}
where
\begin{equation}
T_{\mu\nu}=\Phi_{,(\mu}^{*}\Phi_{,\nu)}-\frac{1}{2}g_{\mu\nu}\left[\Phi_{,\alpha}^{*}\Phi^{,\alpha}+U(|\Phi|^{2})\right]
\end{equation}
and 
\begin{equation}
U(|\Phi|^{2})=\mu^{2}|\Phi|^{2}+\frac{\lambda}{2}|\Phi|^{4},\label{eq:field-pot}
\end{equation}
where $\mu$ is the inverse reduced Compton wavelength of the quantum
of the field and $\lambda$ is the self-interaction coupling constant.
We also define $\Lambda\coloneqq\lambda/(4\pi\mu^{2})$. We choose
units where $\mu=1$, so that the only free parameter of the problem
is $\Lambda$.

\subsection{The ansatz}

We are interested in spherically symmetric static configurations. These can be described by ansatz for the metric (in isotropic coordinates)
\begin{equation}
\d s^{2}=-\e^{2F_{0}(r)}\d t^{2}+\e^{2F_{1}(r)}[\d r^{2}+r^{2}(\d\theta^{2}+\sin^{2}\theta\d\varphi^{2})]\label{eq:metric}
\end{equation}
 and an ansatz for the scalar field
\begin{equation}
\Phi(x^{\mu})=\phi(r)\e^{-\i\omega t},
\end{equation}
where $\omega>0$ is the field's frequency. The time dependence is
necessary to avoid Derrick-type obstructions (also known as virial theorems in this context~\cite{Herdeiro:2022ids}) to the existence of time independent stable scalar lumps~\citep{Derrick1964}. Such
dependence is cancelled in the stress-energy tensor, since it only
depends on the absolute value of the field and its gradients.

\subsection{The boundary conditions}

We have to solve the Einstein-Klein-Gordon equation system, composed
by equations (\ref{eq:einstein-eq}) and (\ref{eq:k-g}). The system
consists of second order ordinary differential equations for the functions
$F_{0},F_{1}$ and $\phi$ along with two boundary conditions for each function. The boundary conditions at the origin to ensure
regularity are given by
\[
\partial_{r}F_{0,1}(0)=0,\quad\partial_{r}\phi(0)=0 ,
\]
whereas asymptotic flatness requires
\[
F_{0,1}(r\to\infty)=\phi(r\to\infty)=0.
\]

\subsection{Solutions}

\begin{figure*}[tpb]
\begin{centering}
\includegraphics{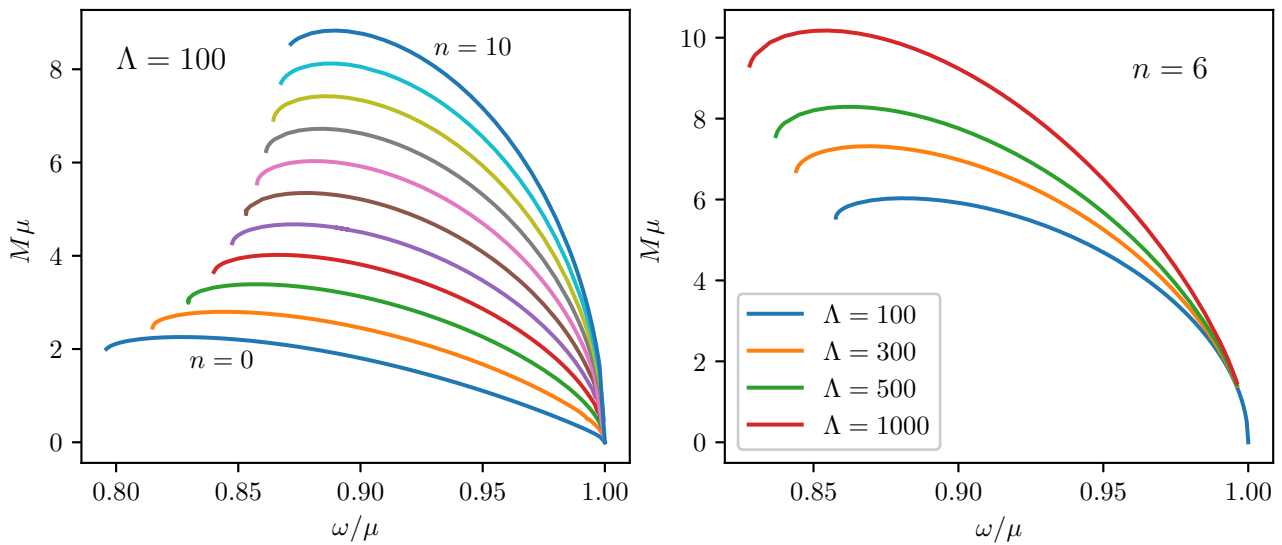}
\par\end{centering}
\caption{\label{fig:ADM-mass-versus}ADM mass versus the frequency for boson
star solutions with fixed $\Lambda$ (left) and fixed $n$ (right).}
\end{figure*}

Solving the equations above will result in a set of infinite solutions
with a different number of radial nodes $n.$ The solutions will only
exist for values of the frequency between $\omega_{{\rm min}}<\omega<\mu$.
The solutions are located on curves such as the ones given in \figref{ADM-mass-versus}, which relate the Arnowitt-Deser-Misner (ADM) mass with the frequency of the stars.
 
In the ground state, solutions located between the maximum of the
ADM mass and $\omega=\mu$ are stable. Analogously, for excited boson
stars the corresponding region will be called candidate stable branch,
following \citep{SanchisGual2022}, but in this case not every solution
in this branch is stable. Although they have been found stable under
infinitesimal perturbations that conserve the total mass and particle
number \citep{Jetzer1992}, for generic perturbations they turn out to be unstable, both in the candidate stable and unstable branch
\citep{Balakrishna1997}. As we will see, however, for large enough values
of the self-interaction coupling constant they can be made stable.

\begin{figure*}[tpb]
\begin{centering}
\includegraphics{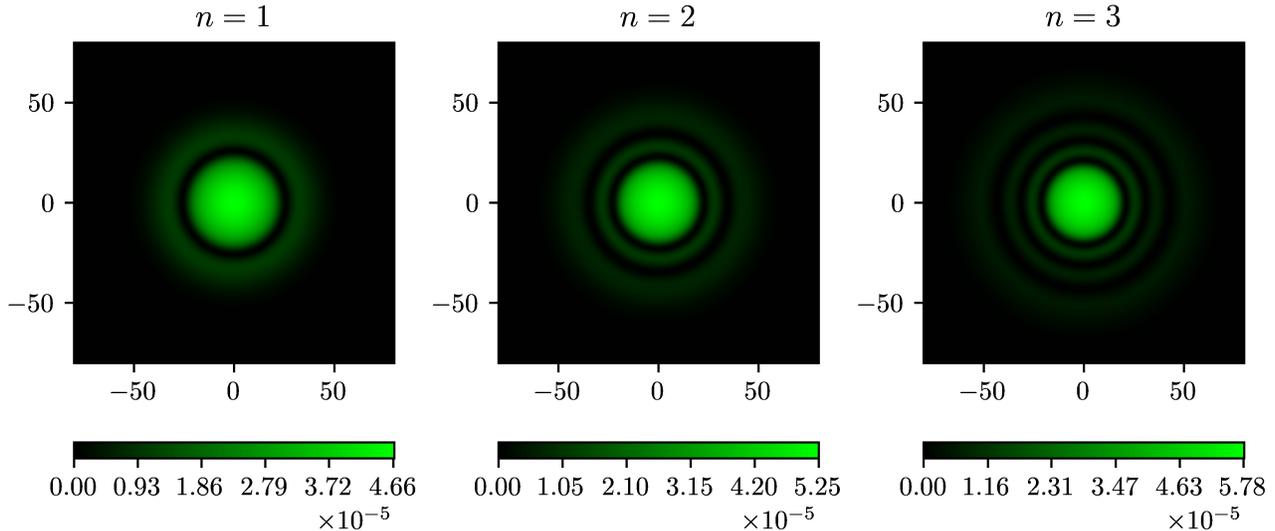}
\par\end{centering}
\caption{\label{fig:Energy-density-of}Energy density  of excited boson stars (omitting
the $\theta$ coordinate) for $\Lambda=500$, $\omega=0.92$, where
$R_{{\rm areal}}=\sqrt{x^{2}+y^{2}}$.}
\end{figure*}

These excited solutions, due to the existence of nodes, will be composed
by a sphere of matter at the centre, surrounded by $n$ shells of
matter. This can be seen in \figref{Energy-density-of}.

\subsection{Mass and compactness}

Knowing the mass of the obtained solutions is important to not only discuss the compactness of the stars, but also to study its evolution
in time, because mass loss during the evolution signals a decay of
the excited state, since states with lower $n$ have less mass than
their more excited counterparts \citep{Friedberg1987}, for the same frequency. As we are
considering an asymptotically flat spacetime, with the vector $\vec{k}=\partial_{t}$
being timelike everywhere, the ADM mass equals the Komar mass evaluated
at infinity, and therefore we have
\begin{equation}
M_\mathrm{ADM}=-\frac{1}{8\pi}\oint\nabla^{\alpha}k^{\beta}\d S_{\alpha\beta}=\int(T-2\tud Ttt)\sqrt{-g}\d r\d\theta\d\phi.\label{eq:komar}
\end{equation}
The mass of these stars increases with the value of $\Lambda$ as
$M\propto\sqrt{\Lambda}$ \citep{Colpi:1986ye} and, for appropriate ranges of the scalar field mass and of the self-interactions coupling, it can be in the solar mass range, supermassive black holes range or dark matter halos range.

Furthermore we can define the compactness of an object as $C\coloneqq M/R$,
where $M$ is its mass and $R$ its areal radius. There is a maximum
compactness of $C=0.5$ which corresponds to an object whose radius
equals its own Schwarzschild radius, therefore becoming a black hole.
Highly compact (but horizonless) stars could, 
in principle, permit the existence of light rings or ISCOs, if their areal radius is smaller than $3M$ or $6M$ respectively,
at least in the case of an exterior Schwarzschild metric. The latter is only an approximation for
spherical bosonic stars; indeed, scalar boson stars (for instance) are modelled by a scalar field which
decays exponentially as $r\to\infty$,
but only reaches zero at infinity. This means that unlike a fluid
star where $T_{\mu\nu}$ vanishes for $r>r_{{\rm surface}}$, there
is no well defined surface radius for boson stars. Still it is possible
to define an effective radius for the star, beyond which we can neglect
the remaining mass of the field, so that the spacetime beyond that
radius is approximately a vacuum spacetime, \emph{i.e.}\ $T_{\mu\nu}\simeq0$.
We shall define such an effective radius as the areal radius containing
99\% of the ADM mass of the spacetime, denoted $R_{99}$~\citep{SanchisGual2022,Schunck:2003kk}.
Furthermore we define compactness of a boson star as 
\begin{equation}
C\coloneqq\frac{M_{99}}{R_{99}},
\end{equation}
where $M_{99}\coloneqq0.99M_{{\rm ADM}}$.

\begin{figure*}[thbp]
\begin{centering}
\includegraphics{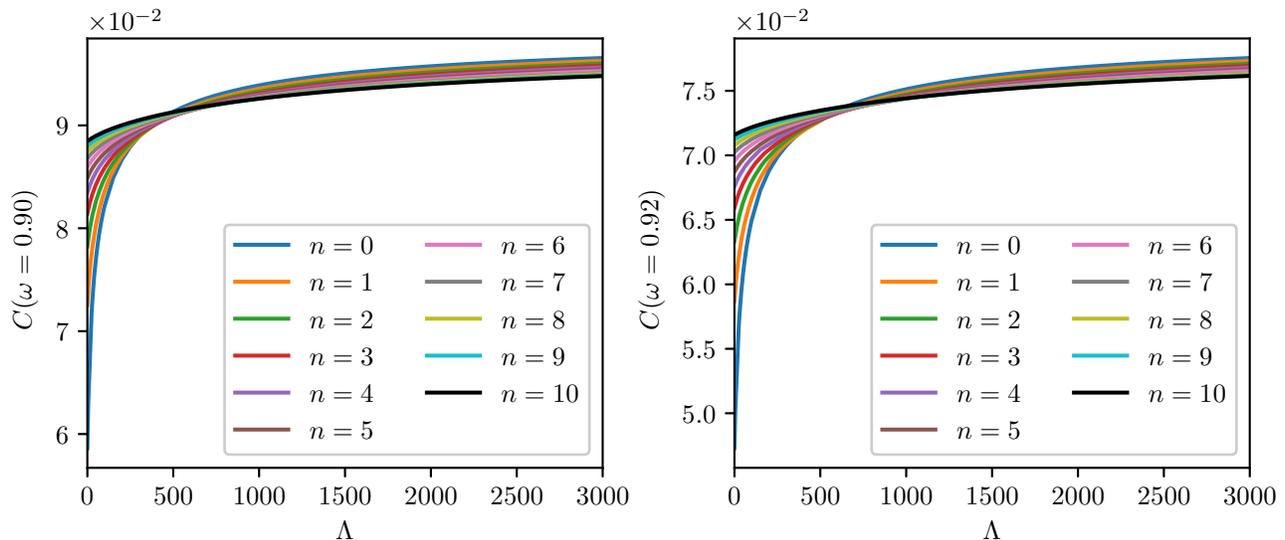}
\par\end{centering}
\caption{\label{fig:compactness}Compactness as a function of $\Lambda$ and $n$ for
$\omega=0.90$ (left) and $\omega=0.92$ (right).}
\end{figure*}
It is expected that the compactness can attain higher values with increasing $\Lambda$
as the case for boson stars in the fundamental state \citep{AmaroSeoane2010},
since the self-interaction is repulsive, and the mass increases with
$\Lambda$. It should also increase as $\omega$ becomes smaller (within
the candidate stable branch) since as $\omega\to1$ we approach the
Newtonian limit.  One might thus expect that, for high enough values of $\Lambda$, the models may start exhibiting ISCOs and unstable circular orbit, as its compactness approaches that of a black hole. It turns out, however, that by increasing $\Lambda$, one seems to
reach a limit of compactness \citep{AmaroSeoane2010}, still well below that of black holes. Thus it is
not possible to turn the star arbitrarily close to a black hole in this way. Our results reinforce this conclusion - figure~\ref{fig:compactness} - where one can see that the value of the compactness seems to asymptote to a value still far from the black hole value, for two fixed (illustrative) values of the frequency. One further notices that for $\Lambda$ close to zero, stars with a larger $n$ are more compact
than stars with smaller $n$. But as $\Lambda\to\infty$ such behaviour
is reversed and stars with smaller $n$ are now more compact than
stars with larger $n$.

\begin{figure}[thbp]
\begin{centering}
\includegraphics{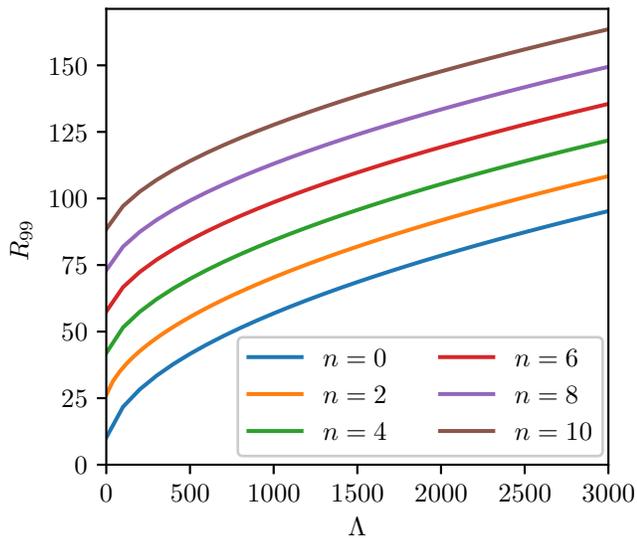}
\par\end{centering}
\caption{\label{fig:effective-radius}Effective radius, $R_{99}$, as a function
of $\Lambda$ and $n$ for $\omega=0.90$.}
\end{figure}
As $n$ increases there is a fairly common asymptotic value for the compactness.
This is a trend observed in most quantities depending on $n$ and
$\Lambda$. The effective radius also increases with $\Lambda$ and
$n$ as seen in \figref{effective-radius},
for boson stars with $\omega=0.90$.  As a consistency check,
it was reported \citep{AmaroSeoane2010} that the maximum compactness
in the stable branch for ground state solutions is $C_{{\rm max}}\approx0.16$.
Since the maximum value of the compactness is for $n=0$, our results
are in agreement with this limit, where $C_{{\rm max}}=0.0965$. In
order to reach this limit one would have to study stars with lower values
of $\omega$ in the candidate stable branches. Moreover, for $n>10$, we would not be able to go past this limit, since the
trend for large $\Lambda$ is that the compactness decreases as $n$
increases.

\section{\label{sec:evolutions}Numerical evolution}

\subsection{BSSN formalism and basic equations}

Using a standard 3+1 spacetime decomposition, a generic spacetime metric can be written in the form
\[
\d s^{2}=g_{\mu\nu}\d x^{\mu}\d x^{\nu}=-\alpha^{2}\d t^{2}+\gamma_{ij}(\d x^{i}+\beta^{i}\d t)(\d x^{j}+\beta^{j}\d t),
\]
where $\alpha$ is the lapse function, $\beta^{i}$ are the shift
functions and $\gamma_{ij}$ is the induced metric on the spatial
hypersurfaces~\citep{Alcubierre2008}.

For the numerical evolutions the spatial metric is further written in the form
\[
\d l^{2}=\e^{4\chi}[a(t,r)\d r^{2}+r^{2}b(t,r)\d\Omega^{2}],
\]
where $\d\Omega^{2}=\d\theta^{2}+\sin^{2}\theta\,\d\varphi^{2}$ and
$a(t,r)$ and $b(t,r)$ are two non-vanishing conformal metric functions,
being related to the physical metric by the conformal decomposition
$\gamma_{ij}=\e^{4\chi}\hat{\gamma}_{ij}$ with $\e^{\chi}=(\gamma/\hat{\gamma})^{1/12}$,
where $\gamma$ and $\hat{\gamma}$ are the determinant of the physical
and conformal 3-metrics respectively. We shall use the Baumgarte-Shapiro-Shibata-Nakamura (BSSN) formulation
\citep{Baumgarte1998,Shibata1995} in spherical coordinates \citep{Alcubierre2011,Montero2012,SanchisGual2015},
which are suited for the problem at hand. The relevant differential
equations for the problem are given in \citep{SanchisGual2022}.

\subsection{Numerical grid and stability}

For the numerical evolutions we use the NADA code, a code for numerical relativity simulations
in 1+1D in spherical symmetry described in \citep{Montero2012,SanchisGual2015,sanchis2015quasistationary,SanchisGual2016,EscorihuelaTomas2017,DiGiovanni2020,DiGiovanni2021}.
The BSSN and Klein-Gordon coupled equations are solved using a second-order
Partially Implicity Runge-Kutta (PIRK) scheme~\citep{Isabel:2012arx,cordero2014partially}. The evolutions are performed
in a logarithmic grid, with a maximum resolution of $\Delta r=0.05$,
a time step of $\Delta t=0.3\Delta r$, the number of radial points
being $n_{r}=50000$ for stars with $n\leq7$ and $n_{r}=80000$ for
the rest, and the outer boundary placed at $r_{{\rm max}}=10000$.
We impose radiative boundary (Sommerfeld) conditions at the outer boundary \citep{Montero2012,Alcubierre:2002kk}.
For the special case of $n=2$, $\omega=0.90$, $\Lambda=125$ the number of radial points
is $n_{r}=100000$ and the outer boundary is placed at $r_{{\rm max}}=20000$, since we extend the evolution to $t=20000$ and we want to avoid reflections from the outer boundary. There are always some reflections, especially from the logarithmic grid when the resolution gets coarse. 

We will not apply any specific perturbations to the stars, since the
numerical truncation error suffices to break the staticity of
the models, in case instabilities are present, triggering their time evolution. Moreover, since the stars are being evolved in a 1+1D code in spherical coordinates under the assumption of spherical symmetry, the perturbations must be spherical. For non-spherical perturbations we need to evolve this models using a 3+1D evolution code.

\section{\label{sec:Results}Dynamical evolution and stability}

We now report the results of the evolution of the static configurations which are solutions
to the Einstein-Klein-Gordon system up to a timescale\footnote{For stars whose composing bosons have the mass of the Higgs ($\unit[125]{GeV}$),
the timescale is around $\unit[5\times10^{-23}]{s}$, with a maximum
mass around $\unit[10^{10}]{kg}$, much smaller than stellar masses.
But for an ultralight boson with a mass of, say $\unit[10^{-10}]{eV}$,
we have a timescale around $\unit[0.066]{s}$ with a maximum mass
of $10M_{\odot}$.} of $t\sim10^{4}$ (where $t$ is measured in units where $\mu=1$).
The evolutions were performed for $n=0,1,\ldots,10$ for a variety
of values of $\Lambda$ in order to find stable boson stars. By a
stable boson stars we mean that within the considered timescale the
initial and final radial profiles coincide, notwithstanding the existence
of oscillations around an equilibrium point between the initial and
final times.

We have made
a detailed study of solutions with $\omega=0.90,0.92$ and we shall
present illustrative cases for one or the other frequency\footnote{These frequencies were chosen since for all values $n=0,...,10$ they belong to the candidate stable branch. The smaller frequency approaches the maximum ADM mass, for $n=10$.}. For the
$n=2$ case we shall present a more detailed study of the solutions, but
such results are similar to the ones we have obtained for other $n>2$.

\subsection{$n=2$ stars}

\subsubsection{Collapse into a black hole}

\begin{figure*}[htbp]
\begin{centering}
\includegraphics{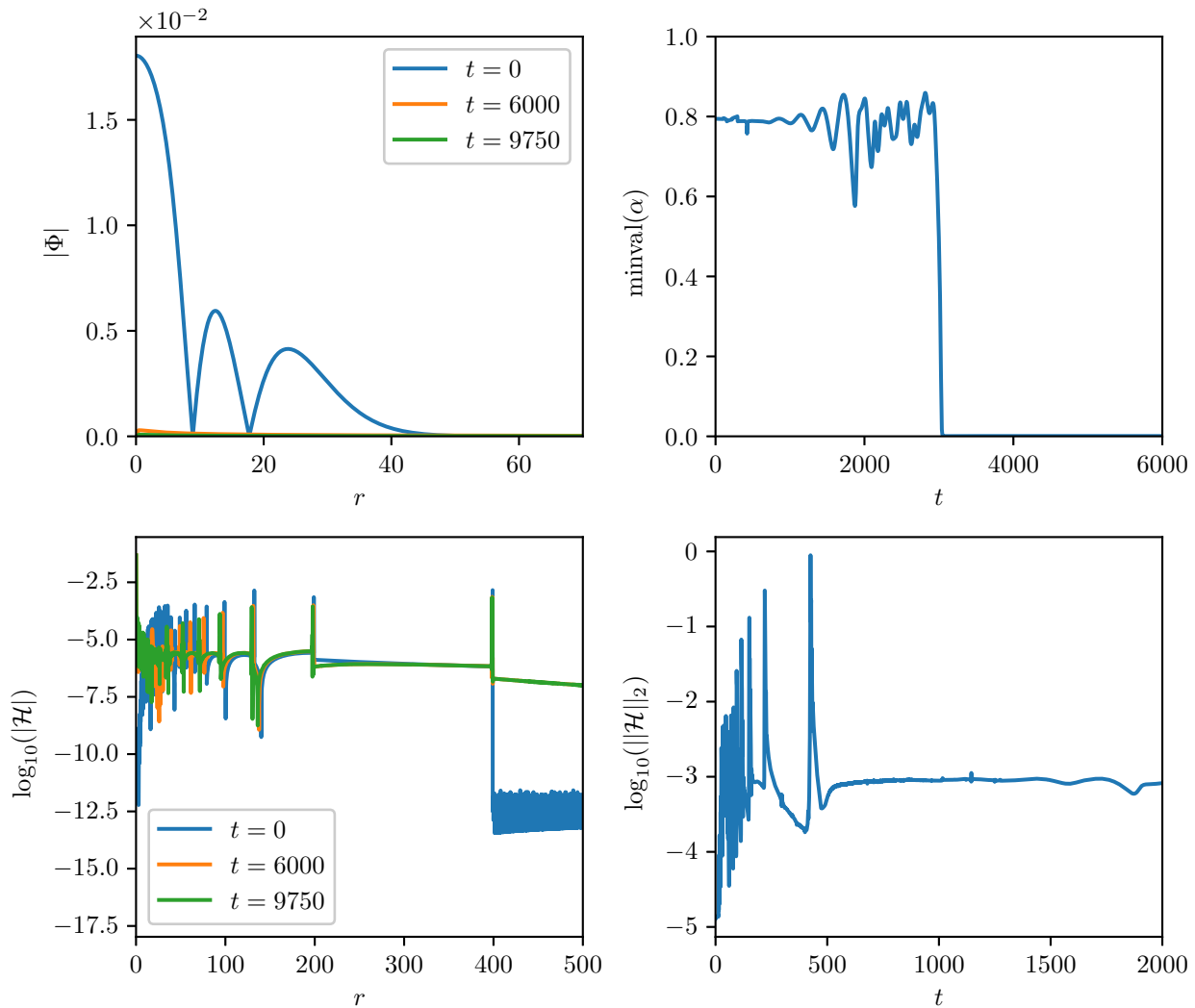}
\par\end{centering}
\caption{\label{fig:Radial-profile-for-black-hole}Radial profile for different
times (top left), minimum value of the lapse (top right), violation
of the Hamiltonian constrain for certain times (bottom left) and the
L2 norm of the Hamiltonian constraint ($||{\cal H}||_{2}$) (bottom
right) for the boson star model $n=2$, $\Lambda=75$ and $\omega=0.92$.
}
\end{figure*}

For small values of $\Lambda$, the star cannot support itself against
its own gravity, and thus it collapses into a black hole -- see \figref{Radial-profile-for-black-hole}.
This is supported by the fact that the minimum value of the
lapse function $\alpha$ drops abruptly after a certain time ($t\approx3000$
in this case), which typically means there was the formation of an apparent
horizon.

\subsubsection{Decay into  a lower state}

\begin{figure*}[htbp]
\begin{centering}
\includegraphics{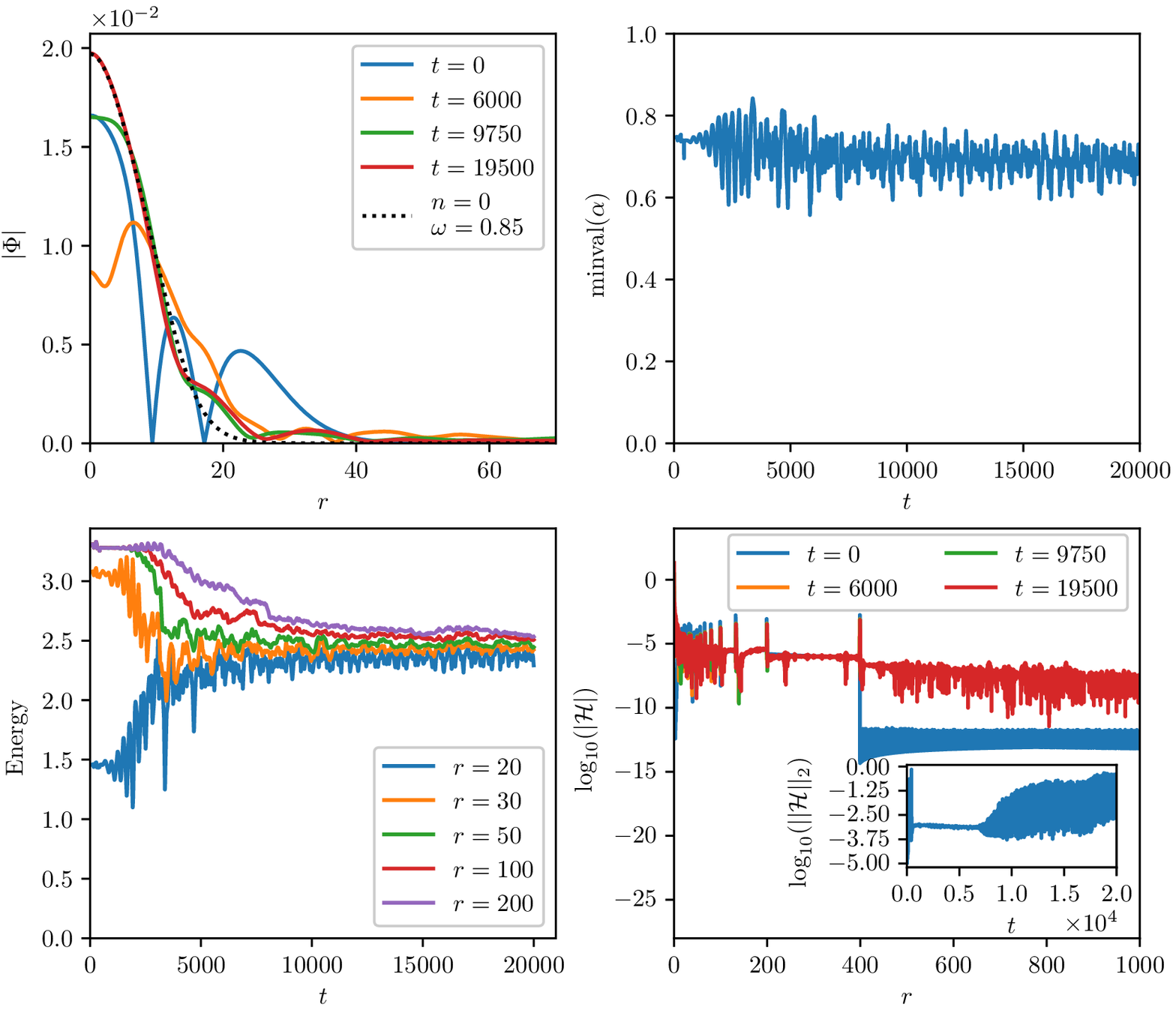}
\par\end{centering}
\caption{\label{fig:Radial-profile-for}Radial profile for different times and radial profile of candidate final state
(top left), minimum value of the lapse (top right), the energy of
the star at different spatial hypersurfaces (bottom left) and violations
of the Hamiltonian constraint for certain times and the L2 norm of
the Hamiltonian constraint (bottom right) for the boson star model
$n=2$, $\Lambda=125$ and $\omega=0.90$.}
\end{figure*}

For intermediate values of $\Lambda$ the stars are still unstable, but with a different outcome -- see \figref{Radial-profile-for}.
The intermediate value of the self-interaction constant $\Lambda$ is still unable to sustain the star, but it manages
to prevent a complete gravitational collapse. It is clear that up to the end of the simulation, the star has not completely
relaxed to the final state, as the ${\rm minval}(\alpha)$ function
shows, but we see in \figref{Radial-profile-for} (bottom left) that, since $r=200$ can be considered as infinity (\emph{cf.}\ \figref{effective-radius}), the star is losing  mass/energy to infinity, which signals a decay.
Furthermore this is accompanied by a redistribution of mass inside the star which is expected since during a decay the star will lose some of its shells, in this case seemingly losing all nodes.
Since the star is not relaxed after $t=10000$, we allowed the evolution to run for $t=20000$.
Even after $t=20000$ the star is not completely relaxed; the plausible conclusion is that it decayed to the $n=0$ state with a different frequency ($\omega=0.85$), since that the energy at $r=200$ is around $E\approx 2.54$ which is the value of the ADM mass of the considered $n=0$ star, which has an $M_{\rm ADM}\approx 2.44$. This suggests that the latter might be the end state of the evolution. Furthermore the central value of the scalar field is consistent with the one from a star with $n=0$, $\Lambda=125$, $\omega=0.85$. For the rest of our models, we were not able to observe a decay of a boson star into a $n\neq0$ state. We evolved several different unstable models that do not collapse into a black hole and found that they all ended up in the fundamental nodeless state. However, we have not explored the entire parameter space and therefore it seems likely that lower excited states can be formed from the decay of unstable excited boson stars.

Let us take this case to discuss the violations of the Hamiltonian
constraint, which provide a diagnostic of the accuracy of our numerical evolution.
Discussing it for this case is sufficient since it has the same features
for all other cases, except when we have a collapse to a black hole,
in which case the violation of the Hamiltonian constraint almost vanishes shortly
after the initial instants. The radial profile of the Hamiltonian constraint in the region where the stars
have support shows that the initial small violations decrease with time. However, the
L2 norm of the Hamiltonian constraint in a larger radial region slowly increases as we get to
the end of the evolution, $t=10000$, but it remains small in the
relevant region where the field is mostly located. It is worth noting
that for all evolutions, the violation of the L2 norm of the Hamiltonian
constraint at later times is larger for $\omega=0.90$ than for $\omega=0.92$,
when it does not collapse into a black hole.

The spikes seen in the violation of the Hamiltonian constraint come from taking second derivatives of the  variables interpolated from the initial data, which has a much lower number of points compared to the grid of the evolution code, specially at $r\geq 50$. This is further supported by the fact that after $r\approx 400$ we have no spikes since the last point of the initial data is around that value.

\begin{figure*}[tpbh]
\begin{centering}
\includegraphics{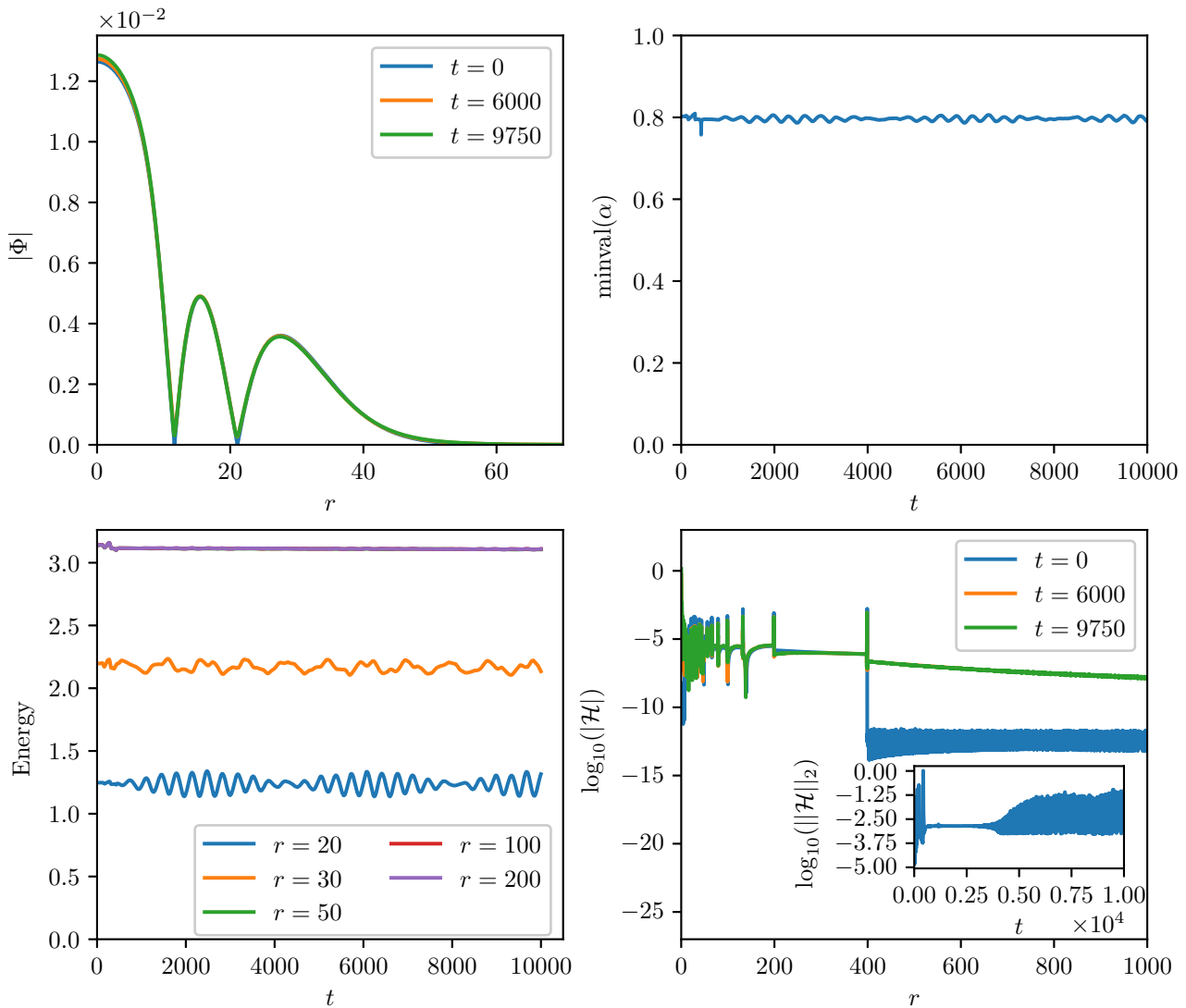}
\par\end{centering}
\caption{\label{fig:Radial-profile-for-1}Radial profile for different times
(top left), minimum value of the lapse (top right), the energy of
the star at different spatial hypersurfaces (bottom left) and violations
of the Hamiltonian constraint for certain times and the L2 norm of
the Hamiltonian constraint (bottom right) for the boson star model
$n=2$, $\Lambda=150$ and $\omega=0.92$.}
\end{figure*}

\subsubsection{Threshold of stability for $n=2$ stars}
For high enough values of $\Lambda$ we found that the stars become
stable. We call such value of $\Lambda$ the threshold of
stability, since for $\Lambda>\Lambda_{{\rm threshold}}$ the stars
are always stable. We have found the threshold of stability for a
star with $\omega=0.90$ to be $\Lambda\simeq 160$ and for a star with
$\omega=0.92$ to be $\Lambda\simeq 150$. As seen in \figref{Radial-profile-for-1}
on the leftmost picture, the star is fully relaxed. There is no mass
loss since the star is stable. For the $n>2$ boson stars studied,
we also found a threshold of stability.

\subsection{Stability of $n>2$ stars}

For the remaining cases of excited boson stars, with $n>2$ we found a similar picture as for the $n=2$ case, with distinct behaviours for low, intermediate and high values of $\Lambda$. Thus, we shall focus now on the thresholds of stability,
that is, the values of $\Lambda$ for which boson star models become
stable. For $\Lambda<\Lambda_{{\rm {\rm threshold}}}$ the stars either
collapse to black holes or into a lower state.

\begin{figure*}[tpbh]
\begin{centering}
\includegraphics{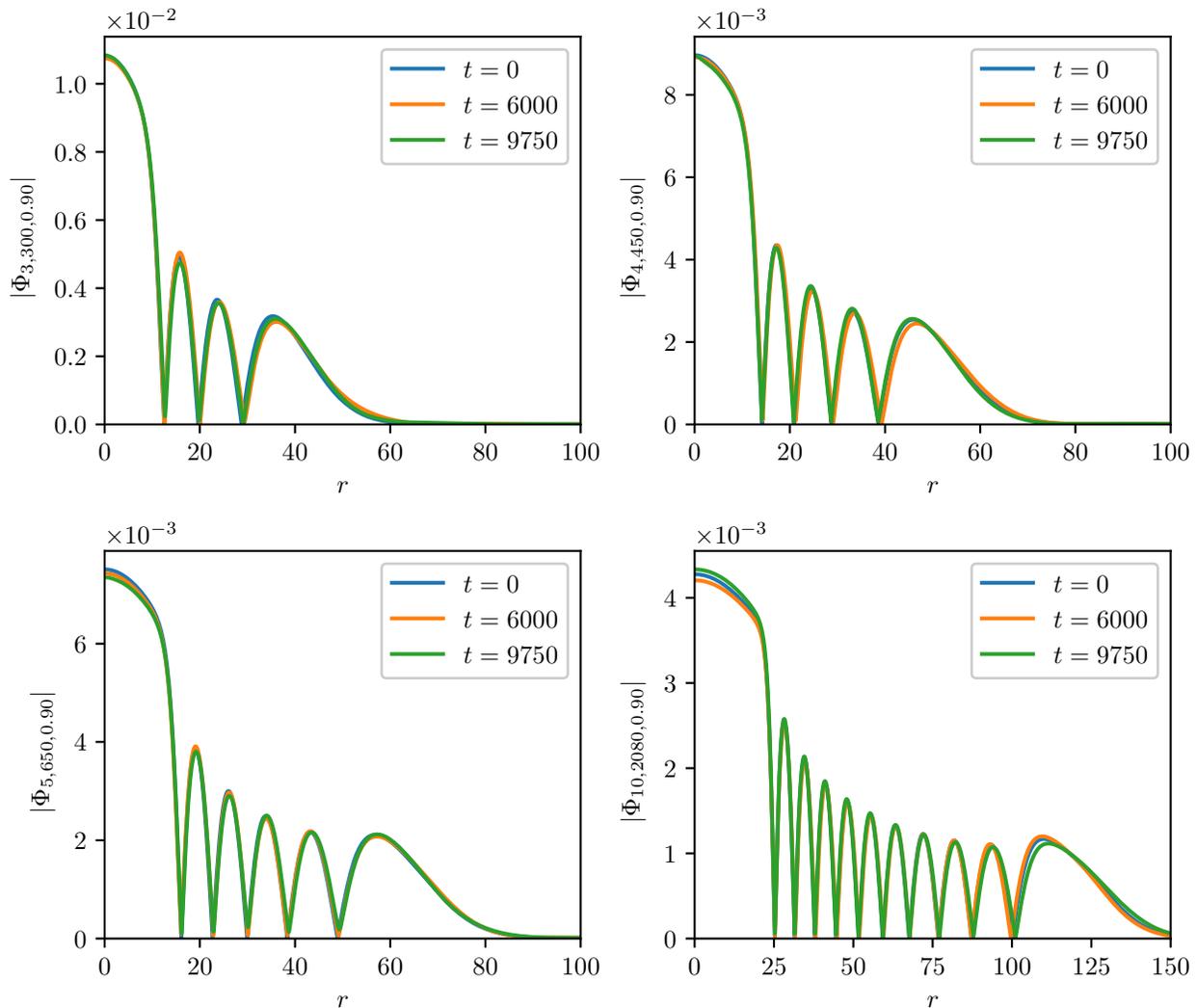}
\par\end{centering}
\caption{\label{fig:radial-profiles}Radial profiles $|\Phi_{n,\Lambda,\omega}|$
for the threshold of stability for several $n$.}
\end{figure*}

The results are quantitatively similar to the previous case, but now
we have an $n$ node star and the threshold of stability happens for
larger values of $\Lambda$. In \figref{radial-profiles} we have
radial profiles for stable models of boson stars for illustrative values
of $n$ and the threshold of stability is shown for two fixed frequencies, in terms of $n$ in table \ref{tab:-for-several}. 
It is worth mentioning that for the $n=0$ stars, we must evolve models
with negative $\Lambda$, to find the threshold of stability, since
they are already stable at $\Lambda=0$. The potential $U(|\Phi|^{2})$
in (\ref{eq:field-pot}) becomes unbounded from below allowing for
infinite negative energies. However, if we stick around the minimum
of the potential we can avoid such problems, at least classically,
where there is no quantum tunnelling. But when $\Lambda$ is negative
enough, we cannot stick around the minimum and such models cannot
even be evolved. For this case the thresholds of stability are $\Lambda=-4$
for $\omega=0.90$ and, at the very least, $\Lambda=-6$ for $\omega=0.92$,
since for $\Lambda<-6$ the evolution crashes.

\begin{table}[H]
\begin{centering}
\begin{tabular}{ccc}
\toprule 
$n$ & $\omega=0.90$ & $\omega=0.92$\tabularnewline
\midrule
\midrule 
0 & $-4$ & $-6$\tabularnewline
\midrule 
1 & $75$ & $74$\tabularnewline
\midrule 
2 & $160$ & $150$\tabularnewline
\midrule 
3 & $300$ & $250$\tabularnewline
\midrule 
4 & $450$ & $390$\tabularnewline
\midrule 
5 & $650$ & $550$\tabularnewline
\midrule 
6 & $850$ & $760$\tabularnewline
\midrule 
7 & $1160$ & $1000$\tabularnewline
\midrule 
8 & $1380$ & $1240$\tabularnewline
\midrule 
9 & $1750$ & $1580$\tabularnewline
\midrule 
10 & $2080$ & $1850$\tabularnewline
\bottomrule
\end{tabular}
\par\end{centering}
\caption{\label{tab:-for-several}$\Lambda_{{\rm threshold}}$ for several
boson star models. For $n=0,1$ the $\Lambda$ of the last unstable
model was taken to be $\Lambda_{n}-1$. For $n>1$ it was taken to
be $\Lambda_{n}-10$.}
\end{table}

\begin{figure}[htbp]
\begin{centering}
\includegraphics{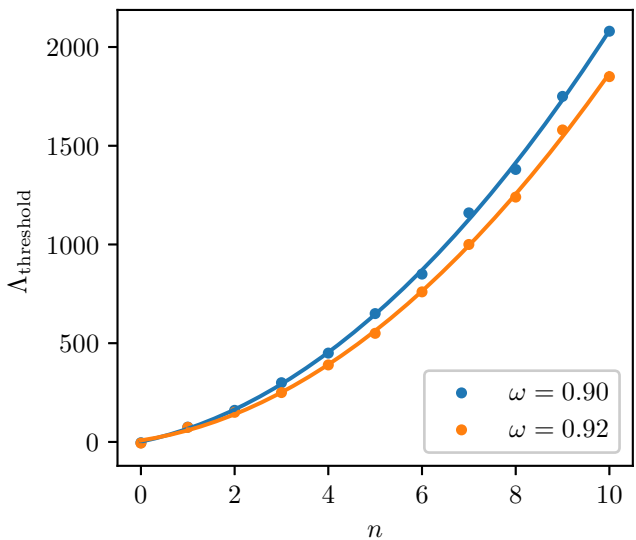}
\par\end{centering}
\caption{\label{fig:stability}The $\Lambda_{{\rm threshold}}$ as a function
of $n$.}
\end{figure}

To conclude, we found that self-interactions can stabilise boson stars
if $\Lambda$ is greater than a certain threshold value. It is worth
remarking that, as  $\Lambda$ grows, stars, initially collapsing
to black holes, begin to avoid the collapse and eventually become
stable when they reach $\Lambda_{{\rm threshold}}$. We found that
as $n$ increases so does $\Lambda_{{\rm threshold}}$, and it appears
that they are related by a quadratic function as seen in \figref{stability}.
This suggests that for all values of $n$ we can stabilise the stars
as long as $\Lambda$ is very large. It would be interesting to establish a mathematical proof of this statement. 

Adjusting a quadratic function to our points we obtain
\begin{align*}
\Lambda & =1.31+49.79n+15.84n^{2}\ \text{for }\omega=0.90,\,R^{2}=0.9997,\\
\Lambda & =7.82+36.16n+14.97n^{2}\ \text{for }\omega=0.92,\,R^{2}=0.9997,
\end{align*}
where $R^2$ is the r-squared of the fit.

\section{\label{sec:ISCO-and-angular}Physical features}

\subsection{Innermost stable circular orbit}

Having established that sufficiently strong self-interactions can stabilise excited scalar
boson stars, it becomes plausible, within the hypothesis that such scalar field models could exist, to consider the potential role of such stars in an astrophysical scenario.
Then, they could have matter surrounding them. If they are
compact enough, tidal disruptions might affect nearby objects leading
to the appearance of an accretion disk, just like in some black holes
and neutron stars. It is thus interesting to see if the models presented
here are compact enough so they support special timelike orbits, like null unstable circular orbits and the ISCO.

\begin{figure}[htbp]
\begin{centering}
\includegraphics{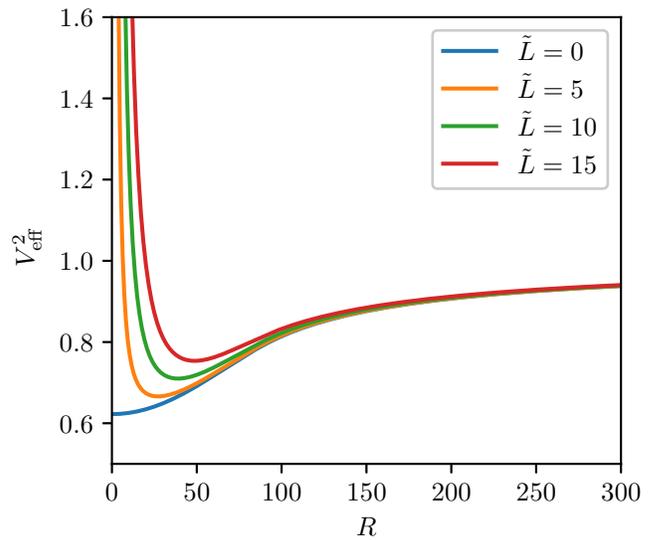}
\par\end{centering}
\caption{\label{fig:Effective-potential-for-1}Effective potential for $n=0,\,\Lambda=3000,\,\omega=0.90$,
the most compact star studied.}
\end{figure}

In order to find the special
orbits for the boson stars, we consider the effective potential
that a particle in a timelike path feels when orbiting a boson star.
We will work with an areal radius, since it is more intuitive
to compare with the well known values for the ISCO and unstable orbits
of the Schwarzschild black hole. The transformation is easily done
knowing that $R_{{\rm areal}}=r_{{\rm iso}}\e^{F_{1}(r_{{\rm iso}})}$.
From now on we denote the areal radial coordinate as $R$.
Then our line element (\ref{eq:metric}) becomes
\begin{equation}
\d s^{2}=-\e^{2F_{0}(R)}\d t^{2}+(1-RF_{1,R})^{2}\d R^{2}+R^{2}\d\Omega^{2}.
\end{equation}
The derivation of the effective potential is a standard textbook
exercise -- see \emph{e.g.}\ \citep{Misner2017}. Consider now a
particle orbiting a boson star. Due to spherical symmetry the orbit can be taken to lie on the equatorial plane and thus $\theta=\pi/2$
and $p_{\theta}=0$ throughout the whole trajectory, where $\vec{p}$
is the particle's four-momentum. Also since we have $\partial_{t}$
and $\partial_{\varphi}$ as Killing vectors due to the static\footnote{Our boson stars are only static after they relax to the final state
since during the evolution they are dynamical.} spherical symmetry of the problem, we can identify $-p_{0}$ as the
energy $E$ and $p_{\varphi}$ as the angular momentum $L$. From
the normalisation of the momentum
\begin{equation}
g_{\mu\nu}p^{\mu}p^{\nu}=-m^{2},
\end{equation}
where $m$ is the mass of the particle, we obtain
\begin{equation}
\underbrace{\e^{2F_{0}}(1-RF_{1,R})^{2}}_{\text{Always positive}}\left(\dd R{\tau}\right)^{2}+\underbrace{\e^{2F_{0}}\left(\frac{\tilde{L}^{2}}{R^{2}}+1\right)}_{V_{{\rm eff}}^{2}(R;\tilde{L})}=\tilde{E}^{2},
\end{equation}
where $\tilde{E}=E/m$, $\tilde{L}=L/m$, $\tau$ is the affine parameter
and $V_{{\rm eff}}^{2}(R;\tilde{L})$ is the effective potential,
plotted in \figref{Effective-potential-for-1}.

The ISCO is found by searching the minimum radius such that
\begin{equation}
\left.\dd{V_{{\rm eff}}^{2}}R\right|_{R_{{\rm min}}(\tilde{L})}=0.
\end{equation}
For our potential such operation must be done numerically. Since the
ISCO is the last stable orbit, we must find out for which values of
$\tilde{L}$, $R_{{\rm min}}$ is the smallest. Again we do
\begin{equation}
\left.\dd{R_{{\rm min}}(\tilde{L})}{\tilde{L}}\right|_{\tilde{L}_{{\rm ISCO}}}=0
\end{equation}
so that we find at which $\tilde{L}$ a particle will stay on the
ISCO and then  $R_{{\rm min}}(\tilde{L}_{{\rm ISCO}})\eqqcolon R_{{\rm ISCO}}$.
There is no ISCO nor unstable circular orbits for the boson stars
studied here since there are stable orbits all the way down to $R\to0$. The
effective potential of the star is qualitatively analogous to the Newtonian potential of an orbiting particle in the Kepler problem, exhibiting  a potential barrier. We conclude that the stars are not compact
enough to support an ISCO. In fact the compactness
of our stars is around $C=R_{99}/M_{99}\sim0.1$. In order to see
an ISCO the effective radius must be at least below $6M$. Even in
the case reported by \citep{AmaroSeoane2010}, $C^{-1}=6.25$, we
still do not have an ISCO. A similar analysis, with adequate adaptations, shows these stars, within the branch studied, have no light rings.

\subsection{Angular velocity of particles}
In~\cite{Olivares:2018abq} (see also \cite{Herdeiro:2021lwl}) it was observed that if the structure of timelike circular orbits around a spherical boson star is such that the angular velocity along the orbits attains a maximum, for an orbit with a non-vanishing radius, then the magneto-rotational instability (MRI) -- a mechanism driving the loss of angular momentum, and therefore driving matter towards the centre of the star -- could cease to be efficient. Thus, such a feature could create a void of matter in the core of the star and an effective shadow, under particular observation conditions. It is thus interesting to examine if such feature could exist for the models of excited boson stars we are considering.

The angular velocity of a particle, as measured by an observer at
infinity, is given by $\d\varphi/\d t$ which is, for an equatorial
orbit,
\begin{equation}
\Omega\coloneqq\dd{\varphi}t=\frac{\d\varphi/\d\tau}{\d t/\d\tau}=\frac{L}{E}\frac{\e^{2F_{0}(R)}}{R^{2}}.
\end{equation}
For circular orbits we know that $\tilde{E}^{2}=V_{{\rm eff}}^{2}$
and that $\dd{}RV_{{\rm eff}}^{2}=0$, since $\dot{R}(\tau)=\ddot{R}(\tau)=0,\forall\tau$.
This allows us to obtain the energy and angular momentum for each
circular orbit. In our case this results in
\begin{align*}
\tilde{E} & =\e^{F_{0}}\left(\frac{-F_{0,R}}{F_{0,R}-1/R}+1\right)^{1/2} \\
\tilde{L} & =\left(\frac{-F_{0,R}}{F_{0,R}/R^{2}-1/R^{3}}\right)^{1/2}
\end{align*}
for circular orbits with radius $R$. Then as a function of the radius
of the orbit we have
\begin{equation}
\Omega(R_{{\rm orbit}})=\sqrt{\frac{\e^{F_{0}}}{R_{{\rm orbit}}}\left.\dd{\,\e^{F_{0}}}R\right|_{R_{{\rm orbit}}}} \,.\label{eq:ang-vel}
\end{equation}

We are interested in studying how the angular velocity changes with $R$ to assess the appearance of a maximum in the
angular velocity for a certain orbit which can give accretion disks
an inner edge even without the presence of an ISCO and resulting in
an effective shadow for the compact object \citep{Olivares:2018abq,Herdeiro:2021lwl},
due to the quenching of the MRI~\citep{Balbus1991}. For mini-boson
stars such features only show up for the unstable branch, having thus
no relevance in an astrophysical scenario. For a free
vector field instead of a scalar one, such features do show up \citep{Herdeiro:2021lwl}
in the stable branch. Our excited boson stars, being composed by multiple
spherical shells surrounding one another, and having a self-interaction,
might behave differently from mini-boson stars.

\begin{figure}[htbp]
\begin{centering}
\includegraphics{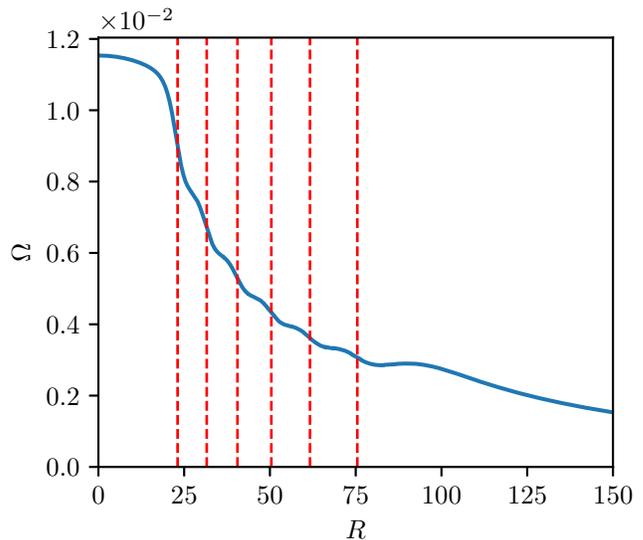}
\par\end{centering}
\caption{\label{fig:Angular-velocity-as}Angular velocity as a function of
$R_{{\rm orbit}}$ for $n=6$, $\Lambda=800$, $\omega=0.92$. The
vertical lines show the location of the nodes.}
\end{figure}

The existence of nodes in our boson star models, has an influence
in the angular velocity of a particle, depending on whether its orbit
is located around a node or not. In fact the existence of nodes causes
the appearance of plateaus in the angular velocity, located in between
the nodes of the radial function as seen in \figref{Angular-velocity-as}.
As $R\to0$, the angular velocity stops increasing and a new plateau
starts to appear. This holds even in the mini-boson star case, where the plateau
is very small, being much more pronounced for large values of
$\Lambda$. The reason is that as one approaches the origin, the matter contained inside a sphere of radius R also vanishes, and in a way that so does the variation of the orbital
velocity. 

In between the nodes we have a similar situation since,
at the nodes we have a vacuum, which is followed by a shell of matter.
Since part of the mass of the star is enclosed in a 2-sphere of $R<R_{{\rm node}}$,
near the node the mass function is approximately constant which
allows the particle to increase its angular velocity approximately\footnote{This holds in a Newtonian approximation which does not take
into account the pressure of the scalar field.} as $\sim\sqrt{M/R^{3}}$. But similarly to the $R\to0$ case, as we
cross the shell of mass between the nodes, the enclosed mass in a
sphere of radius $R$ will decrease and $\Omega(R)$ will change as
$\sim\sqrt{M(R)/R^{3}}$, which decreases the slope of the angular
velocity curve. Also the radius of the star increases with $\Lambda$
(although the mass also increases), which explains why the region
of the plateaus is wider.

\begin{figure*}[tpbh]
\begin{centering}
\begin{center}
\includegraphics{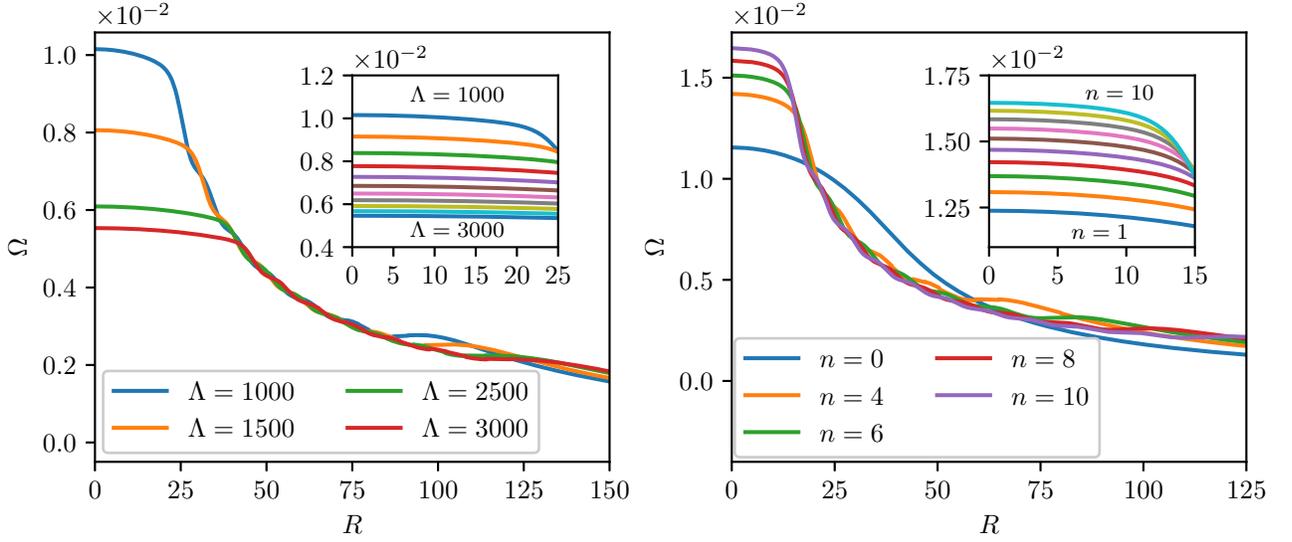}
\par\end{center}
\par\end{centering}
\caption{\label{fig:Angular-velocity-as-1}Angular velocity as a function of
$R_{{\rm orbit}}$ for $n=6$, $\omega=0.92$ for several $\Lambda$
(left). Angular velocity as a function of $R_{{\rm orbit}}$ for $\Lambda=300$,
$\omega=0.92$ for several $n$ (right). Same qualitative behaviour
for $\omega=0.90$.}
\end{figure*}

It is also worth noting that, as the value of $\Lambda$ increases,
the angular velocity near the centre seems to approach an asymptotic value. The same also happens when $\Lambda$ is fixed and $n$ varies, again suggesting that for very large $\Lambda$ the number of nodes seems
not to matter. These features can be seen in \figref{Angular-velocity-as-1}.

Furthermore we report that for every $n\geq3$ and for all $\Lambda$, for both of our studied 
frequencies, there is a maximum in the angular velocity, just before the angular velocity decays in a
Keplerian fashion. That maximum is very small, questioning if it is a mere numerical artifact. 
However performing several tests, such as increasing the number of points, obtaining the solutions in
other coordinate systems, changing the interpolation, we were not able to get rid of this unexpected 
behaviour. Assuming it is physical, raises the question as to why it occurs and, moroever, 
only for $n>2$, since no qualitative 
difference between $n=2$ and $n=3$ was to be expected.

\subsection{Galactic rotation curves}

The existence of nodes in boson stars has important consequences for
the rotational velocity of matter around them. Facing these boson stars as dark matter, suggests comparing these rotational velocities to galactic rotation curves.

As it is well known, the rotational
velocity of stars in galaxies do not follow the expected Keplerian
behaviour $v_{{\rm rot}}\simeq\sqrt{GM(R)/R}$, the further we get
from the centre of the galaxy, where $M(R)$ is the mass of the visible
or luminous matter enclosed in a sphere of radius $R$ (for a review of dark matter in galaxies and its interaction with the baryonic matter see \citep{Salucci:2018hqu}). Since we have
a high density of matter distributed around the centre, at first $v_{{\rm rot}}$
increases, but once we are past that region, $v_{{\rm rot}}$ should
decrease as $\propto\sqrt{1/R}$. Instead it has been famously found that the rotational
velocity does not decrease in this way well past the region which contains the luminous
matter. One proposed explanation is that galaxies are surrounded by
dark matter halos, which could hypothetically be galactic scale boson stars.

\begin{figure*}[tpbh]
\begin{centering}
\includegraphics{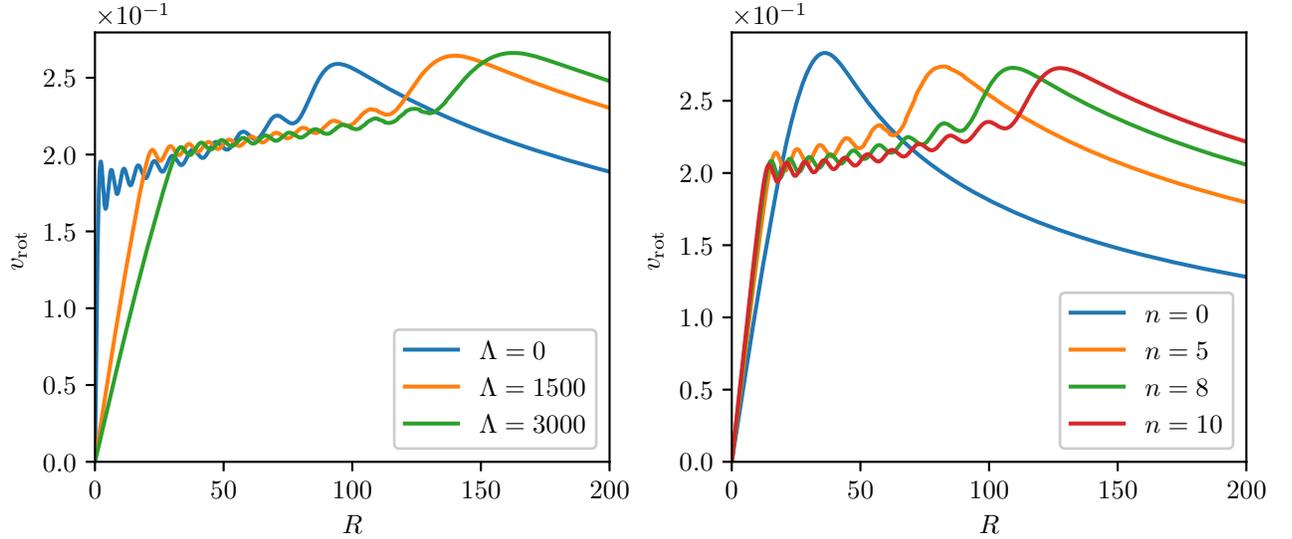}
\par\end{centering}
\caption{\label{fig:rot-velocity}Rotational velocity as a function of $R_{{\rm orbit}}$
for $n=10$, $\omega=0.92$ for several $\Lambda$ (left). Angular
velocity as a function of $R_{{\rm orbit}}$ for $\Lambda=500$, $\omega=0.92$
for several $n$ (right). Same qualitative behaviour for $\omega=0.90$.}
\end{figure*}
The existence of plateaus in the $\Omega(R)$ plots hints at an increase of the rotational velocity in certain regions.
The rotational velocity is defined as $v_{{\rm rot}}\coloneqq R\Omega$
and a plot versus the radius of orbit is given in \figref{rot-velocity}.
Excited boson stars with self interactions
had already been proposed as candidates to galaxy halos in \citep{Lee:1995af}. This analysis is, however, different from ours, since therein the Newtonian rotational
velocity is used, which does not take into account the pressure which is non-negligible
\citep{Schunck1998}.

We can divide the plots in \figref{rot-velocity}
into three regions. First, near $R=0$ there is a steep linear increase
in $v_{{\rm rot}}$. This is expected since galaxies have a distribution
of visible mass at the centre which decreases as $R\to R_{{\rm 1st\,node}}$.
This is also what is expected assuming a Keplerian $v_{{\rm rot}}$
for galaxies, since the observable mass is also mainly located at
the centre. However beyond the visible mass, $v_{{\rm rot}}$ should
decay as $\sqrt{1/R}$, which is not the case, as experimental evidence
shows that it increases linearly. Then there is a second region where
the rotational velocity increases almost linearly albeit with some
small oscillations, caused by the nodes. As the numbers of nodes increases
the size of this region also increases, so stars with many nodes could
explain the increases of rotational velocities in galaxies. In the
last region we recover the expected $\sqrt{1/R}$ decay, which would
correspond to a particle outside the galactic halo.

For a certain number of nodes, the increase of $\Lambda$ only increases
the effective radius of the star and thus the rotational velocity
of the stars, although the qualitative features are the same for all
$\Lambda$, they happen at larger radius as $\Lambda$ increases.
The putative astrophysical importance of $\Lambda$ here is that a sufficiently
large value can stabilise a boson star with a large number of nodes.

The foregoing discussion is, however, only a statement of principle. If we try to apply the studied models to real 
galaxies, the rotational velocity does not drop below $\sim \unit[10^4]{km/s}$, whereas in realistic 
galaxies, velocities are $\sim \unit[10^2]{km/s}$. The studied models are way too compact for applications to real galaxies and were studied
within the context of stellar compact objects. If boson stars are to be the galactic 
dark matter halos, much more dilute boson stars, approaching the Newtonian limit, would be the correct framework.

\section{Discussion and conclusion}

It is well known that scalar boson stars in the ground state are stable
against perturbations even in the absence of self-interactions, in some regions of the parameter space.
In \citep{SanchisGual2022}, however, it was shown that self interactions can stabilise excited boson stars with one radial node, $n=1$, for certain
values of $\Lambda$. Using the same approach we have shown that such
is also the case for stars with a number of nodes between $2\leq n\leq10$,
as long as $\Lambda$ is greater than a certain threshold value. It
is also expected that for sufficiently high enough values of $\Lambda$,
stars with $n>10$ should also be stable, since there is no reason for 
any qualitative difference between $n<10$ and $n>10$ stars. It would be interesting to attempt a mathematical proof of this statement. Furthermore, we have obtained an empirical quadratic
relationship between $\Lambda_{{\rm threshold}}$ and $n$. We, however, leave the question of stability with respect to more general (non-spherical) perturbations for a future work, since it requires these models to be evolved in a 3+1D code.

Although differing from mini-boson stars in many ways, these self-interacting
excited boson stars are not compact enough so that features like ISCOs or light
rings can be observed. We are focusing on candidate stable
branch solutions, since only those can be made dynamically stable.
It has been reported before that mini-boson stars can be be
highly compact in the unstable branch, which, however, has limited physical significance. The compactness was also found to be increasing with $\Lambda$
and it tends to an asymptotic value as $\Lambda\to\infty$, where
stars with lower $n$ are more compact than stars with larger $n$.
But for large $n$ the compactness of the stars seems to become (roughly) independent
of $n$.

The angular velocity of test particles in circular orbits in boson
stars spacetimes was also studied and it was found that the angular
velocity as a function of $r$ has several plateaus, that is, tends
to become constant, near the location of the nodes. The existence
of plateaus, hints at peculiar angular velocity profiles, as found in \figref{rot-velocity},
where there is a region far from the centre where we witness an almost
linear increase of the rotational velocity, which is qualitatively
similar to what has been observed in galactic rotation curves. This suggesting fitting such models (after the threshold of stability)  to real
galactic rotation curves. No such comparison with data was made, since this required stars with a frequency $\omega$ close to $1$, the Newtonian limit.
Solutions with $\omega=0.9999$ were obtained but even in that case the 
velocities were ten times larger than the velocities of realistic galaxies. One would
need to obtain solutions in the Newtonian limit (solving the Schrödinger-Poisson
equation), as done in \cite{Sin:1992bg} to fit with real galactic data, and then repeat the stability analysis. This is an interesting research direction for future work. We remark that it was previously discussed that stars with nodes could have a rotational velocity profile similar to those of galaxies; the potential stability of these objects, however, was not discussed. Our work shows that the stability issue could be solved within the paradigm of self-interactions.

Since these excited stars can, in principle, be realisable
in an astrophysical context (even the dynamical formation of such
objects can be possible as discussed in \citep{SanchisGual2022} for
$n=1$ stars), one could also consider simulations of the collision of these objects, to assess their gravitational wave signals and compare them with real data from LIGO-Virgo-KAGRA detections.

\begin{acknowledgments}
This work is supported  by the  Center for Research and Development in Mathematics and Applications (CIDMA)
through the Portuguese Foundation for Science and Technology (FCT -- Fundaç\~ao para a Ci\^encia e a 
Tecnologia), references  UIDB/04106/2020 and UIDP/04106/2020.  
The authors acknowledge support  from the projects CERN/FIS-PAR/0027/2019, PTDC/FIS-AST/3041/2020, 
CERN/FIS-PAR/0024/2021 and 2022.04560.PTDC\@.  
This work has further been supported by  the  European  Union's  Horizon  2020  research  and  innovation
(RISE) programme H2020-MSCA-RISE-2017 Grant No.~FunFiCO-777740 and by the European Horizon Europe staff
exchange (SE) programme HORIZON-MSCA-2021-SE-01 Grant No.~NewFunFiCO-101086251. M.B. is supported by the FCT
grant GDG-2/2021/BI/CIDMA and 2022.09704.BD\@. This work is also supported by the Spanish Agencia Estatal de Investigaci\'on (Grant PID2021-125485NB-C21). NSG is supported by the Spanish Ministerio de Universidades, through a María Zambrano grant (ZA21-031) with reference UP2021-044, funded within the European Union-Next Generation EU. Computations have been performed at the Argus
and Blafis cluster at the U.~Aveiro and at the Navigator Cluster at the LCA in U.~Coimbra through project 2021.09676.CPCA\@.
\end{acknowledgments}

\appendix
\section{Numerical convergence}

\begin{figure}[tpbh]
    \centering
    \includegraphics{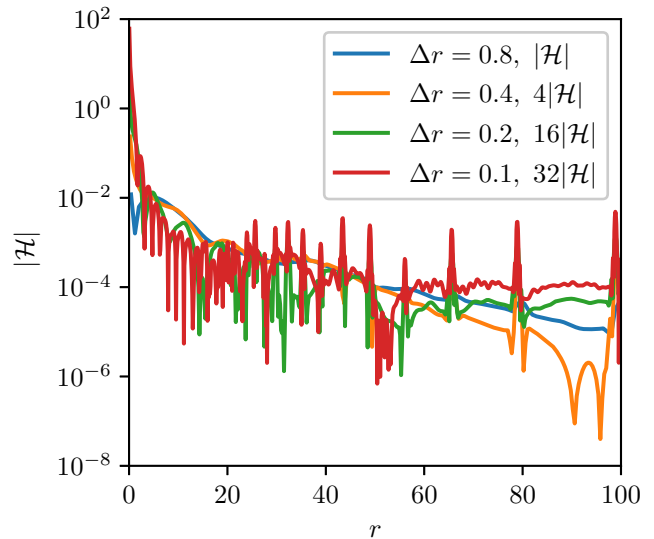}
    \caption{Absolute value of the Hamiltonian constraint at $t=1200$ for different resolutions for a boson star with $n=2$, $\omega=0.92$, $\Lambda=250$.}
    \label{fig:ham-const-conv}
\end{figure}
\begin{figure}%[tpbh]
    \centering
    \includegraphics{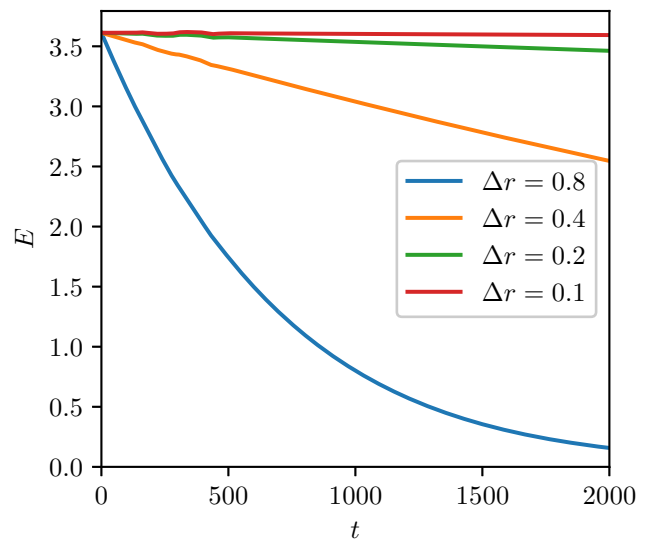}
    \caption{Evolution of the total mass/energy for different resolutions.}
    \label{fig:mass-conv}
\end{figure}
\begin{figure*}
    \centering
    \includegraphics{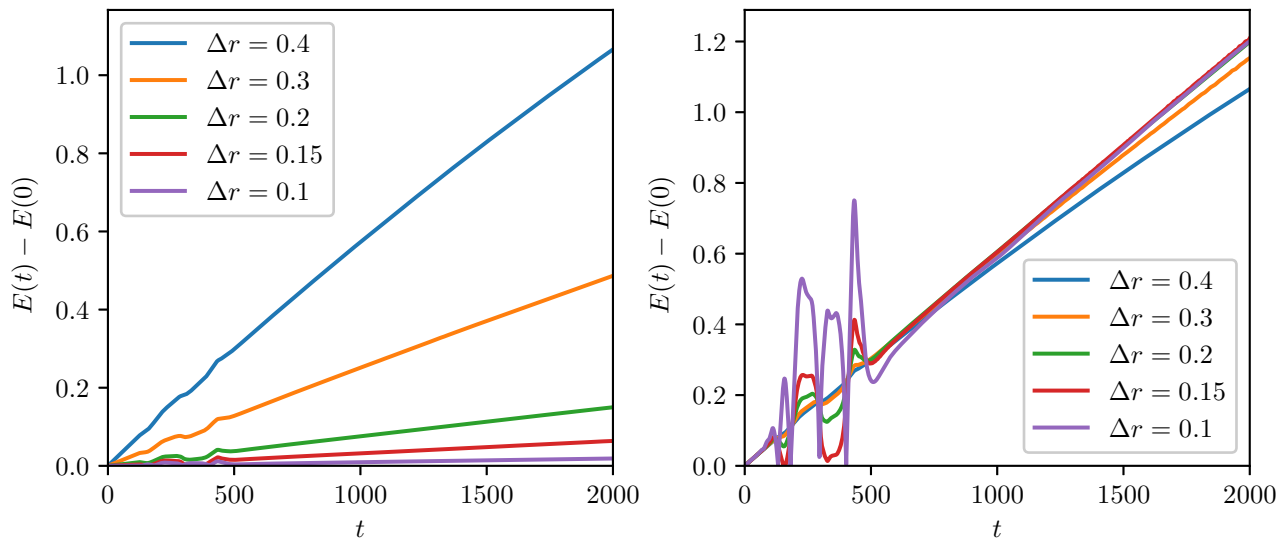}
    \caption{Difference between total mass at $t=1200$ and total initial mass (left) and re-scaled functions to third-order convergence (right).}
    \label{fig:res-mass-conv}
\end{figure*}

In order to assess the quality of our numerical simulations we perform a convergence test, consisting in comparing different quantities from various grid resolutions and see if the results converge to the expected value.
In order to perform our numerical evolutions we imported initial data into the code which was then interpolated to the evolution grid. We consider only numerical error coming from the finite difference operations, which dominates the error if we use resolutions coarser than the initial data resolution.
In \figref{ham-const-conv} we show the absolute value of the Hamiltonian constraint for four different resolutions at an instant of time $t=1200$ for a stable boson star with $n=2$, $\omega=0.92$ and $\Lambda=250$.
We find the expected second-order convergence, since the PIRK (Partially Implicitly Runge-Kutta) time integrator is second order~\cite{Montero2012}, for the first three resolutions ($\Delta r=0.8$, $\Delta r=0.4$, $\Delta r=0.2$). However, as the resolution is further increased, the convergence order is reduced to between first and second order. This is due to the fact that for high resolutions we are not improving anything compared to the initial grid.

We can also compare the drift of the evolution of the total mass as in \figref{mass-conv}. Since we consider a stable model the total mass should be constant throughout the evolution.
However, due to numerical error, the mass decreases with time and for low resolutions the numerical solutions are not good enough. Taking the deviation from the initial value of the mass $E(t=0)$
we find that the order of convergence is 3, due to the fourth-order interpolatation, the second-order PIRK and the fourth-order finite differencing. The scaled functions can be seen in \figref{res-mass-conv}.

\bibliographystyle{apsrev4-1}

\end{document}